\documentclass[11pt,a4paper]{article}
\usepackage{slashed}
\usepackage{jheppub}

\usepackage{subfig}
\usepackage{bm}
\usepackage{color}
\usepackage{float}
\usepackage[usenames]{xcolor}
\usepackage{hyperref}
\usepackage{siunitx}
\hypersetup{colorlinks=true,urlcolor=blue,linkcolor=magenta,citecolor=blue,filecolor=blue}
\usepackage[normalem]{ulem}
\usepackage{array}
\usepackage{hyperref}
\usepackage{booktabs}
\usepackage{blindtext}
\usepackage{upgreek}
\usepackage{subfloat}
\usepackage[normalem]{ulem}
\newcommand{\stkout}[1]

\allowdisplaybreaks

\usepackage{float}
\usepackage[scr=rsfs]{mathalfa}
\usepackage{bigints}
\usepackage[english]{babel}
\usepackage{float}
\usepackage{braket}

\newcommand*{\Tr}{\operatorname{Tr}}

\newcommand{\diff}[1]{\text{d}#1}
\newcommand{\dd}{\text{d}}

\newcommand{\id}{\mathbb{I}}
\newcommand{\dv}[2]{\frac{\text{d}#1}{\text{d}#2}}

\begin{document}

\title{Self-dual instantons and gravitating dyons in non-Abelian ModMax theory}

\author[a,b]{Fabrizio Canfora,}
\author[c]{Crist\'obal Corral,}
\author[c]{Borja Diez,}
\author[d]{Luis Guajardo,}
\author[e]{Julio Oliva}

\affiliation[a]{Centro de Estudios Científicos (CECs), Casilla 1469, Valdivia, Chile}
\affiliation[b]{Facultad de Ingenier\'ia, Arquitectura y Diseño, Universidad San Sebastian \\ General Lagos 1163, Valdivia 5110683, Chile}
\affiliation[c]{Departamento de Ciencias, Facultad de Artes Liberales, Universidad Adolfo Ibáñez, Avenida Padre Hurtado 750, 2562340, Viña del Mar, Chile}
 \affiliation[d]{Instituto de Matem\'aticas (INSTMAT), Universidad de Talca, Casilla 747, Talca 3460000, Chile.}
 \affiliation[e]{Departamento de F\'isica, Universidad de Concepci\'on, Casilla, 160-C, Concepci\'on, Chile.}

\emailAdd{fabrizio.canfora@uss.cl}
\emailAdd{cristobal.corral@uai.cl}
\emailAdd{borjadiez1014@gmail.com}
\emailAdd{luis.guajardo.r@gmail.com}
\emailAdd{juoliva@udec.cl}

\abstract{Motivated by the recent interest in conformal and duality invariant nonlinear electrodynamics, we study the non-Abelian extension of ModMax electrodynamics. The theory is parameterized by a single dimensionless constant, and it is continuously connected to Yang-Mills theory in its vanishing limit. We show that the theory admits (anti-)self-dual instantons, despite the additional nonlinearities that characterize the non-Abelian ModMax theory. For $SU(2)$, we construct the generalization of the BPST instanton and extend this solution to Euclidean de Sitter and anti-de Sitter backgrounds. In the latter case, the Chern-Pontryagin index depends on the instanton size since the configuration is not a pure gauge at infinity; a property already pointed out in Yang-Mills on negative-curvature backgrounds by Callan and Wilczek. We compute the contribution of the latter to the spectrum of the Dirac operator at the boundary, which is crucial for determining the non-local contributions to the Dirac index. Then, we show that the ansatz constructed with 't Hooft symbols accommodates multi-instantons in the non-Abelian ModMax theory. The system of (anti-)self-dual equations reduces to a single nonlinear equation, which can be perturbatively solved order by order in the parameter that controls the nonlinearity. Following such a strategy, we provide a formal solution for the $N$-instanton configuration to first order in the expansion. Then, we couple non-Abelian ModMax theory to gravity with a conformally coupled scalar field and construct new gravitating solutions that describe Euclidean wormholes and other smooth configurations with secondary hair. }

\maketitle

\section{Introduction\label{sec:intro}}

The interest in nonlinear extensions of electrodynamics dates back to the early days of the past century, when it was realized that a point-like model for the electron leads to a divergent self-energy within Maxwell's theory. Attempts to remove this pathology led to the well-known Born-Infeld theory~\cite{Born:1934gh}. This theory is governed by a dimensionful coupling that measures the degree of nonlinearity, and it converges continuously to Maxwell theory in the limit when this coupling vanishes. This type of deformation to classical electrodynamics emerges as the effective action controlling the dynamics of gauge fields living on D-branes~\cite{Fradkin:1985qd,Leigh:1989jq}. In quantum electrodynamics, on the other hand, loop corrections introduced by massive fermions can be integrated out, leading to the Euler-Heisenberg Lagrangian that supplements Maxwell action with a nonlinear term suppressed by one over the fermion mass to the fourth power~\cite{Heisenberg:1936nmg} (for a review of the topic see~\cite{Sorokin:2021tge} and references therein). Then, it became clear that nonlinear corrections generically appear in any field theory when quantum effects are taken into account. 

Recently, it has been realized that it is possible to deform Maxwell theory with a dimensionless coupling $\gamma$, maintaining, at the same time, the electromagnetic duality of the dynamics and the invariance of the action under the conformal group~\cite{Bandos:2020jsw}. The latter implies that if such a continuous deformation exists, it must be controlled by a dimensionless coupling. Such a theory, discovered in Ref.~\cite{Bandos:2020jsw}, was dubbed ModMax. 
It was later rederived in Ref.~\cite{Kosyakov:2020wxv} using the Gaillard-Zumino-Gibbons-Rasheed formalism for duality rotations in nonlinear electrodynamics~\cite{Gibbons:1995cv,Gaillard:1981rj,Gibbons:1995ap,Gaillard:1997zr,Gaillard:1997rt}, and more recently in Ref.~\cite{Ayon-Beato:2024vph} employing the Salazar-Garc\'ia-Pleba\'nski formulation~\cite{Salazar:1987ap}.

Since ModMax emerges naturally as the most general nonlinear electrodynamics possessing the same features as Maxwell theory, its physical consequences have been thoroughly explored over the last few years. The study of the spectrum of propagating degrees of freedom was done in Ref.~\cite{Bandos:2020jsw}, where it was also shown that there is birrefringence in vacuum. Supersymmetric extensions were constructed in~\cite{Bandos:2021rqy}, while in~\cite{Flores-Alfonso:2020euz,Amirabi:2020mzv,Kubiznak:2022vft,Barrientos:2022bzm,Barrientos:2024umq} it was shown that its coupling to gravity leads to interesting curved spacetimes as accelerating black holes, multi-black hole solutions~\cite{Bokulic:2025usc}, Euclidean solitons~\cite{Corral:2025npd}, NUTs and Bolts~\cite{BallonBordo:2020jtw,Flores-Alfonso:2020nnd}, Eguchi-Hanson instantons~\cite{Colipi-Marchant:2023awk}, and hairy black holes~\cite{Zhang:2021qga,Ayon-Beato:2024vph}. The Galilean and Carrollian limits have been taken and explored in~\cite{Banerjee:2022sza,Correa:2024qej} (see also~\cite{Chen:2024vho}). Recently, holographic aspects of ModMax theory have been explored in Ref.~\cite{Barrientos:2025rde}, and their connection to $T\bar{T}$-deformations in conformal field theories was studied in Refs.~\cite{Rodriguez:2021tcz,Tempo:2022ndz,Ferko:2022cix,Babaei-Aghbolagh:2022leo,Babaei-Aghbolagh:2022uij,Ferko:2023wyi,Babaei-Aghbolagh:2024hti,Babaei-Aghbolagh:2024uqp}. 

In Yang-Mills theory, instantons are Euclidean, stationary, and regular (anti-)self-dual solutions that minimize the action on a given topological sector~\cite{Belavin:1975fg}. They also control quantum effects as tunneling between topologically inequivalent vacua, and they are fundamental in a non-perturbative formulation of the quantum theory~\cite{tHooft:1976rip,tHooft:1976snw}. However, due to topological reasons, Euclidean $U(1)$ gauge theories do not have finite action, localized, regular instantonic solutions on flat space, a feature that is cured by the non-Abelian extensions. Given the relevance of the instantonic configurations, it is natural to ask whether they exist in a non-Abelian extension of the ModMax theory, where the notion of self-duality is natural, and how they are deformed with respect to the instantons in Yang-Mills. This is the aim of the present work.

We start by considering a sensible one-parameter family of non-Abelian extension of ModMax theory in Sec.~\ref{sec:Non-Abelian ModMax theory}. The theory is parameterized by a single dimensionless constant $\gamma$, which reduces to Yang-Mills theory in the $\gamma \to 0$ limit. We demonstrate that the theory admits (anti-)self-dual configurations with a vanishing energy-momentum tensor, despite the inherent nonlinearities of the ModMax model. Focusing on the $SU(2)$ case, in Sec.~\ref{sec: Instantons on constant curvature spaces} we construct the generalized BPST instanton on flat space and extend the solution to Euclidean de Sitter ($\mathbb{S}^4)$ and anti-de Sitter ($\mathbb{H}^4)$ backgrounds. In the latter case, the Chern-Pontryagin index is shown to depend on the instanton size, since the configuration is not a pure gauge at infinity; a feature already reported by Callan and Wilczek in Yang-Mills theory on negative-curvature backgrounds~\cite{Callan:1989em}. Section~\ref{sec:Multi-instantons} considers the 't~Hooft ansatz, which accommodates multi-instantons in Yang-Mills theory~\cite{tHooft:1976snw}. Indeed, we show that such a choice is also very useful in the non-Abelian ModMax context. Remarkably, the resulting (anti-)self-dual system reduces to a single nonlinear equation, which we solve perturbatively in the dimensionless parameter. To first order, we obtain a formal solution for the $N$-instanton configuration. In Sec.~\ref{sec:Dirac}, we compute the spectrum of the Dirac operator on the boundary of Euclidean AdS space, which is crucial for obtaining the non-local contributions to the Dirac index in the presence of non-Abelian fields. In Sec.~\ref{sec:Self-gravitating}, we couple the Euclidean non-Abelian ModMax theory to Einstein gravity in the presence of a conformally-coupled scalar field and construct new gravitating solutions, including Euclidean wormholes and gravitational instantons with a secondary scalar hair. We conclude with comments in Sec.~\ref{sec:Discussion}.

\section{Non-Abelian ModMax theory}\label{sec:Non-Abelian ModMax theory}

Let us consider the $SU(2)$ as the simplest non-Abelian gauge group for extending ModMax theory. We start by setting up our conventions for the generators of $SU(2)$, which are given by $t_a=-\frac{i}{2}\tau_a$, with $\tau_a$ being the Pauli matrices, which satisfy $[\tau_a,\tau_b]=2i\epsilon_{abc}\tau^c$ and $\{\tau_a,\tau_b\}=2\delta_{ab}\id$, where $\delta_{ab}$ and $\epsilon_{abc}$ correspond to two invariant tensors of the $SU(2)$ group. These relations imply that the generators satisfy
\begin{equation}
	[t_a,t_b]=\epsilon_{abc}t^c\qquad \text{and}\qquad \Tr(t_at_b)=-\frac{1}{2}\delta_{ab}\,.
\end{equation}
The non-Abelian gauge potential $1$-form in the adjoint representation of $SU(2)$ is defined as $A_\mu=A_\mu^{a}t_a$, with its associated non-Abelian field strength given by
\begin{equation}\label{Fmunudef}
	F_{\mu\nu}=\partial_\mu A_\nu-\partial_\mu A_\nu + [A_\mu,A_\nu ]\,.
\end{equation}
The non-Abelian ModMax action is constructed out of two gauge invariants, which are even and odd under parity transformations; they are
\begin{equation}\label{XandY}
	X=\frac{1}{4}\Tr (F_{\mu\nu}F^{\mu\nu})\qquad \text{and}\qquad Y=\frac{1}{4}\Tr(\tilde{F}_{\mu\nu}F^{\mu\nu})\,,
\end{equation}
respectively, where $\tilde{F}_{\mu\nu}=\frac{1}{2}\epsilon_{\mu\nu\lambda\rho}F^{\lambda\rho}$ is the dual of the field strenght, and $\epsilon_{\mu\nu\lambda\rho}$ denotes the anti-symmetric Levi-Civita tensor. 

The extension of ModMax electrodynamics to the $SU(2)$ case has been studied in Ref.~\cite{Cirilo-Lombardo:2023poc}. It was shown that the latter can be obtained from the secular equation that arises from the four eigenvectors associated with the four main directions in the Newman-Penrose classification~\cite{Newman:1961qr}. In Euclidean signature, the resulting conformally-invariant action is given by\footnote{Here, we will work mainly in Euclidean signature unless otherwise stated. The Lorentzian version of Eq.~\eqref{nAMM} is obtained by simultaneously changing the global sign of the action and performing $Y^2\to-Y^2$.}
\begin{equation}\label{nAMM}
	I_{\rm NAM}[g,A]=-\int_\mathcal{M}\dd^4x\sqrt{|g|}\left(X\cosh\gamma-\sqrt{X^2-Y^2}\,\sinh\gamma\right)\,.
\end{equation}
The field equations associated with the non-Abelian field $A_\mu$ are obtained by performing stationary variations of the action with respect to it, giving
\begin{equation}\label{eom-A}
	\nabla_\mu P^{\mu\nu}+[A_\mu,P^{\mu\nu}]=0\,,
\end{equation}
where we have defined the constitutive tensor as
\begin{equation}\label{P-tensor}
	P_{\mu\nu}=\left(\cosh\gamma-\frac{X\sinh\gamma}{\sqrt{X^2-Y^2}}\right)F_{\mu\nu}+\frac{Y\sinh\gamma}{\sqrt{X^2-Y^2}}\tilde{F}_{\mu\nu}\,.
\end{equation}
Notice that (anti-)self-dual configurations in Yang-Mills theory are singular here, as they satisfy $X=\pm Y$. Nevertheless, as we will see below, there is a generalized (anti-)self-duality condition in non-Abelian ModMax theory that allows one to obtain regular configurations. On the other hand, the Bianchi identity in this case reads
\begin{equation}\label{Bianchi-indentity}
	\nabla_\mu\tilde{F}^{\mu\nu}+[A_\mu,\tilde{F}^{\mu\nu}]=0\,.
\end{equation}

It is worth mentioning that Gibbons and Rasheed studied a global $SO(2)$ electromagnetic duality invariance of general nonlinear electrodynamics in Ref.~\cite{Gibbons:1995cv}. They concluded that the condition $\tilde{F}_{\mu\nu}F^{\mu\nu}=\tilde{P}_{\mu\nu}P^{\mu\nu}$ ensures that the theory is invariant under duality rotations. One can check that the non-Abelian generalization of the Gibbons-Rasheed condition, that is,
\begin{align}
\Tr(\tilde{F}_{\mu\nu}F^{\mu\nu})=\Tr(\tilde{P}_{\mu\nu}P^{\mu\nu})\,,
\end{align}
is satisfied if the definition in Eq.~\eqref{P-tensor} is assumed. However, Eqs.~\eqref{eom-A} and~\eqref{Bianchi-indentity} are not invariant under global $SO(2)$ duality transformations, as they both depend explicitly on the non-Abelian gauge connection. Hence, due to the nonlinear nature of the field strength, there is no local transformation of the non-Abelian gauge connection that implements a continuous $SO(2)$ rotation.

On the other hand, conformal invariance implies that the stress-energy tensor must be traceless. In the case of the non-Abelian theory, the latter is given by
\begin{equation}\label{Tmunu}
T_{\mu\nu}^{(A)}=\Tr(P_{(\mu}{}^{\lambda}F_{\nu)\lambda})-g_{\mu\nu}\left(X\cosh\gamma-\sqrt{X^2-Y^2}\sinh\gamma\right)\,.
\end{equation}
Then, one can check easily that the condition $g^{\mu\nu}T_{\mu\nu}=0$ is satisfied off-shell by virtue of the identity 
\begin{align}\label{PF}
\Tr(P^{\mu\nu}F_{\mu\nu})=4\left(X\cosh\gamma - \sqrt{X^2-Y^2}\,\sinh\gamma\right)\,.    
\end{align}
Additionally, in Euclidean signature, the stress-energy tensor~\eqref{Tmunu} can be rewritten as
\begin{equation}\label{TmunuSD}
     T_{\mu\nu}^{(A)} = \frac{1}{2}\Tr\left[\left(\tilde{P}_{\mu\lambda} + F_{\mu\lambda} \right)\left(P_{\nu\rho} - \tilde{F}_{\nu\rho} \right)\right]\,g^{\lambda\rho}\,.
\end{equation}
In this form, it is clear that it vanishes identically for configurations satisfying the generalized (anti-)self-dual conditions
\begin{equation}\label{BPS-cond}
	P_{\mu\nu}=\pm \tilde{F}_{\mu\nu}\,.
\end{equation}
This is a first-order nonlinear differential equation that defines instantons in nonlinear conformal electrodynamics. Field configurations satisfying this condition solve the non-Abelian field equations~\eqref{eom-A} identically by virtue of the Bianchi identity. We will explore these configurations in the following.

\section{(Anti-)self-dual instanton on constant curvature spaces}\label{sec: Instantons on constant curvature spaces}

In order to look for instantonic solutions on four-dimensional Euclidean backgrounds of constant curvature, we define the left-invariant forms of the $SU(2)$ group, as
\begin{subequations}
	\begin{align}
  \sigma_1&=\cos\psi\dd\vartheta + \sin\vartheta\sin\psi\dd\varphi\,,\\
    \sigma_2&=-\sin\psi\dd\vartheta + \sin\vartheta\cos\psi\dd\varphi\,,\\
    \sigma_3&=\dd\psi + \cos\vartheta\dd\varphi\,,
\end{align}
\end{subequations}
where $(\vartheta,\varphi,\psi)$ are Euler angles with $\vartheta\in[0,\pi]$, $\varphi\in[0,2\pi)$, and $\psi\in[0,4\pi]$. These $1$-forms satisfy the Maurer-Cartan equation $\dd\sigma_a+\tfrac{1}{2}\epsilon_{abc}\sigma^b\wedge\sigma^c=0$. The metric of a constant-curvature space can be parametrized using a radial coordinate $r$ and the $SU(2)$ left-invariant one-forms $\sigma_i$, such that the line element takes the form
\begin{equation}\label{dsCCS}
	\dd s^2=\dd r^2+\frac{\sin^2(\sqrt{k}r)}{4k}\,(\sigma_1^2+\sigma_2^2+\sigma_3^2)\,.
\end{equation}
Here, $k=\pm1,0$ describes the Euclidean de Sitter ($\mathbb{S}^4$), anti-de Sitter ($\mathbb{H}^4$), and flat ($\mathbb{R}^4$) spaces, respectively. The range of the radial coordinate is $r\in[0,\pi]$ for $k=1$ and $r\in \mathbb{R}_{\geq0}$ otherwise. The Riemann curvature of the metric~\eqref{dsCCS} is $R_{\ \mu\nu}^{ \lambda\rho}=k\,\delta_{\mu\nu}^{\lambda\rho}$, where the generalized Kronecker delta is defined as $\delta^{\mu_1\ldots\mu_p}_{\nu_1\ldots\nu_p}=p!\,\delta^{[\mu_1}_{[\nu_1}\dots\delta^{\mu_p]}_{\nu_p]}$ .

For the non-Abelian potential, we choose an ansatz aligned along the left-invariant forms of $SU(2)$, that is,
\begin{equation}\label{A}
	A=A_\mu^{a}t_a\dd x^\mu=a(r)\sigma^{a}t_a\,.
\end{equation}
The non-Abelian field strength can be read off from its definition in Eq.~\eqref{Fmunudef}, giving
\begin{equation}\label{F-k}
	F=\frac{1}{2}F_{\mu\nu}\dd x^\mu\wedge\dd x^\nu=\left(a'\dd r\wedge\sigma^{i}+
	\frac{a(a-1)}{2}\epsilon^{i}_{~jk}\sigma^j\wedge\sigma^k\right)t_i\,,
\end{equation}
where prime denotes differentiation with respect to the radial coordinate $r$. Written in this form, the electric and magnetic components of the non-Abelian field strength are evident. 

In order to solve the field equations~\eqref{eom-A}, we use the generalized (anti-)self-duality conditions in Eq.~\eqref{BPS-cond}. First, we evaluate the $SU(2)$ gauge invariants of Eq.~\eqref{XandY} on the ansatz for the non-Abelian potential defined in Eq.~\eqref{A}, giving 
	\begin{align}\label{XY-k}
  X&=-\frac{3ka'^2}{\sin^2(\sqrt{k}r)}-\frac{12k^2a^2(a-1)^2}{\sin^4(\sqrt{k}r)} & &\mbox{and} &
  Y&=-\frac{12k^{3/2}a'a(a-1)}{\sin^3(\sqrt{k}r)}\,.
\end{align}
Notice that the flat-space limit of these invariants ($k\to0$) is smooth, allowing us to treat the three geometries on the same footing. Thus, the constitutive relation $2$-form in Eq.~\eqref{P-tensor} takes the form
\begin{align}
	P=\frac{1}{2}P_{\mu\nu}\dd x^\mu\wedge\dd x^\nu&=\left(e^{-\gamma}a'\dd r\wedge \sigma^{i}+\frac{a(a-1)e^\gamma}{2}\epsilon^{i}_{~jk}\sigma^j\wedge\sigma^k\right)t_i\,,
\end{align}
which is very similar to the non-Abelian field strength in Eq.~\eqref{F-k}, provided a suitable rescaling of their electric and magnetic parts. Then, by replacing these results into the BPS condition in Eq~\eqref{BPS-cond}, we obtain a remarkably simple single first-order differential equation, that is,
\begin{align}\label{dadr-k}
	\dv{a}{r}=\pm\frac{2e^\gamma \sqrt{k}a(a-1)}{\sin(\sqrt{k}r)}\,.
\end{align}

This equation can be solved analytically for $k=\pm 1$ and $k=0$. However, as the equation bifurcates for different values of $k$, first, one needs to take the limit $k\to\pm1$ or $k\to0$ before integrating Eq.~\eqref{dadr-k}. This procedure yields
\begin{align}\label{asol-k}
    a_\pm(r) &= 
    \begin{cases}
        \displaystyle \frac{1}{1+\rho_\pm 
        \tan\!\left(\tfrac{\sqrt{k}r}{2}\right)^{\pm 2e^\gamma}}
        & \text{for } k=\pm 1, \\[1.2em]
        \displaystyle \frac{1}{1+\rho_\pm r^{\pm 2e^\gamma}} 
        & \text{for } k=0 \,.
    \end{cases}
\end{align}
Here, $\rho_\pm$ denotes an integration constant related to the instanton size, while the $\pm$ sign corresponds to the self-dual and anti-self-dual branches of the solution, respectively. Notice that, in the case of the flat Euclidean background with $\gamma\to 0$, the solution reduces to the standard Belavin-Polyakov-Schwartz-Tyupkin (BPST) instanton~\cite{Belavin:1975fg}. For the other constant curvature backgrounds, namely $k=\pm 1$, the solutions reduce to those obtained by Ivanova-Lechtenfeld-Popov (ILP)~\cite{Ivanova:2017wun} in the Yang-Mills limit. 

It is worth mentioning that, in the cases $k=0$ and $k=1$, the profile in~\eqref{asol-k} is pure gauge at the origin and asymptotically zero, whereas the anti-instanton exhibits the opposite behavior: it is zero at the origin and asymptotically pure gauge. However, in the case $k=-1$, the (anti-)instanton is (zero) pure gauge at the origin, but still depends on the integration constant asymptotically, that is,
\begin{equation}\label{apminfty}
	\lim_{r\to \infty}a_\pm (r):= a_{0\pm} =\frac{1}{1+\rho_\pm }\,,\qquad \text{for }k=-1\,.
\end{equation}
This behavior will have consequences when computing global quantities, such as topological invariants or the Euclidean action, as we will see next. Additionally, all these solutions are regular, and non-trivial provided that $\rho_\pm\neq0$, as can be verified from the $SU(2)$ invariants, which we summarize in Table~\ref{tab:SU2inv-asol-k}.
\begin{table}[h]\centering
\renewcommand{\arraystretch}{1.8}
\setlength{\tabcolsep}{12pt}     
\begin{tabular}{| c | c | c |}
\hline
    & $k=\pm1$ & $k = 0$ \\
    \hline
    $X_\pm$ & 
    $\displaystyle -\frac{12\rho_\pm^2(1+e^{2\gamma})\,
    \tan\!\left(\tfrac{\sqrt{k}r}{2}\right)^{\pm 4e^\gamma}}
    {\sin(\sqrt{k}r)^4\,\left(1+\rho_\pm \tan\!\left(\tfrac{\sqrt{k}r}{2}\right)^{\pm 2e^\gamma}\right)^4}$ &
    $\displaystyle -\frac{12\rho_\pm^2(1+e^{2\gamma})\,r^{\pm 4e^\gamma-4}}
    {(1+\rho_\pm r^{\pm 2e^\gamma})^4}$ 
    \\
    \hline
    $Y_\pm$ & 
    $\displaystyle \mp \frac{24e^\gamma \rho_\pm^2
    \tan\!\left(\tfrac{\sqrt{k}r}{2}\right)^{\pm 4e^\gamma}}
    {\sin(\sqrt{k}r)^4\,\left(1+\rho_\pm \tan\!\left(\tfrac{\sqrt{k}r}{2}\right)^{\pm 2e^\gamma}\right)^4}$ &
    $\displaystyle \mp \frac{24e^\gamma\rho_\pm^2 \, r^{\pm 4e^\gamma-4}}
    {(1+\rho_\pm r^{\pm 2e^\gamma})^4}$ 
    \\ 
    \hline
\end{tabular}
\caption{$SU(2)$ invariants for the (anti-)self-dual solution in Eq.~\eqref{asol-k}. Here, $X_\pm$ and $Y_\pm$ denote evaluation of the gauge invariants in Eq.~\eqref{XandY} on the (anti-)self dual configurations presented in Eq.~\eqref{asol-k}.}
\label{tab:SU2inv-asol-k}
\end{table}

Indeed, when $k\to1$ or $k\to0$, the solution belongs to the same homotopy class as the BPST instanton of Yang-Mills theory~\cite{Belavin:1975fg}, as it can be seen by computing its Chern-Pontryagin index, that is,
\begin{align}\label{Chern-Pontryagin}
    C_2[A] = -\frac{1}{16\pi^2}\int_{\mathcal{M}}\dd^4x\sqrt{|g|}\,\Tr(\tilde{F}_{\mu\nu}F^{\mu\nu}) = \pm1\,,
\end{align}
for self-dual and anti-self-dual instantons, respectively. On the other hand, the partition function can be obtained from the Euclidean on-shell action which, to first order in the saddle-point approximation, is given by $\ln\mathcal{Z}\approx -I_E$. The latter can be computed directly by using the identity~\eqref{PF} and the (anti-)self-dual condition $P_{\mu\nu}=\pm\tilde{F}_{\mu\nu}$, giving
\begin{align}\label{INAM-on-shell}
    I_{\rm NAM}\Big|_{\rm on-shell} &= -\frac{1}{4}\int_{\mathcal{M}}\dd^4x\sqrt{|g|}\,\Tr(P_{\mu\nu}F^{\mu\nu}) = \pm 4\pi^2 \,C_2[A]\,.
\end{align}
Using the Chern-Pontryagin index for the (anti-)self-dual configurations in Eq.~\eqref{Chern-Pontryagin}, we find that the Euclidean on-shell action for these instantons on Euclidean de-Sitter and flat space is $4\pi^2$.

The case $k=-1$ is somewhat subtle, as the solution in Eq.~\eqref{asol-k} is not asymptotically pure gauge. Even more, since Euclidean AdS has a conformal boundary, one needs to include the contribution of boundary terms to correctly account for topological invariants of the Pontryagin class. Indeed, as explained in Ref.~\cite{Callan:1989em}, due to the nontrivial asymptotic behavior of non-Abelian gauge fields in negative-curvature spaces, the Chern-Pontryagin index of $SU(2)$ does not need to be an integer. In this case, we find that
\begin{equation}\label{CP-Ad4}
	C_2[A]_+=\frac{(3+\rho_+ )\rho_+^2}{(1+\rho_+)^3}\,,\qquad C_2[A]_-=-\frac{1+3\rho_-}{(1+\rho_- )^3}\,,
\end{equation}
for self-dual and anti-self-dual solutions, respectively, while the Euclidean on-shell action is exactly that given in Eq.~\eqref{INAM-on-shell}. To account for the correct value of topological invariants of the Pontryagin class in the presence of boundaries, one needs to take into account the spectral asymmetry of the Dirac operator at the boundary~\cite{Atiyah:1975jf,Atiyah:1976jg,Atiyah:1976qjr}. We will compute the spectrum of the Dirac operator in Sec.~\ref{sec:Dirac}. Notice that, in the large (small) instanton-size limit, the $SU(2)$ Chern-Pontryagin index of the self-dual solution is one (zero), while for the anti-self-dual instanton is zero (minus one).

\section{Beyond the BPST instanton}\label{sec:Multi-instantons}

Constructing multi-instantons in the non-Abelian ModMax theory is a rather difficult task, due to its highly nonlinear structure on top of the already nonlinear Yang-Mills theory. Nevertheless, one of the main characteristics of ModMax theory is precisely scale invariance. Thus, at least at the classical level, there is no characteristic scale. This fact hints that it is possible to construct multi-instanton configurations similar to those of Yang-Mills, where the positions and relative distances between the single-centered instantons are arbitrary, partly due to the absence of a characteristic length scale.

To take first steps toward this goal, we consider flat Euclidean space $\mathbb{R}^4$ as the background, parametrized by Cartesian coordinates $x=(x^i,\tau)$ where $\tau$ is the Euclidean time and $i=1,2,3$ are $SO(3)$ indices. Then, we introduce the self-dual and anti-self-dual 't~Hooft symbols~\cite{tHooft:1976snw}
\begin{subequations}
\begin{align}
	\eta_{a\mu\nu}&:=\epsilon_{a\mu\nu}+\delta_{a\mu}\delta_{\nu 4}-\delta_{a\nu}\delta_{4\mu}  \,,\\
	\bar{\eta}_{a\mu\nu}&:=\epsilon_{a\mu\nu}-\delta_{a\mu}\delta_{\nu 4}+\delta_{a\nu}\delta_{4\mu}\,, 
\end{align}
\end{subequations}
respectively, satisfying
\begin{equation}
	\eta_{a\mu\nu}=\frac{1}{2}\epsilon_{\mu\nu\rho\sigma}\eta_{a}^{~\rho\sigma}\;\;\;\;\; \mbox{and}\;\;\;\;\; \bar{\eta}_{a\mu\nu}=-\frac{1}{2}\epsilon_{\mu\nu\rho\sigma}\bar{\eta}_a^{~\rho\sigma}\,.
\end{equation}
More details about the properties of these symbols can be found in Ref.~\cite{tHooft:1976snw}. These objects are constant tensors that encode the local decomposition of the four-dimensional rotation group according to $SO(4)\simeq SU(2)\times SU(2)$. In this way, they provide a mapping between $SO(3)$ indices of constant Euclidean-time hypersurfaces and the $SU(2)$ indices of the internal group. 

Then, we construct the following ansatz for the $SU(2)$ gauge field
\begin{equation}
	A_\mu = -\bar{\eta}_{\mu\nu}\partial^\nu \log \Phi(x) \,,\label{tHansatz}
\end{equation}
where $\Phi=\Phi(x)$ is a function of the coordinates, with no particular symmetry assumed, and $\bar{\eta}_{\mu\nu}:=\bar{\eta}_{a\mu\nu}\,t^a$. In the case of Yang-Mills theory, substituting this ansatz into the self-dual condition $F_{\mu\nu}=\tilde{F}_{\mu\nu}$ reduces them to $\nabla^2 \Phi(x)=0$, which is solved by~\cite{tHooft:1976snw}
\begin{equation}\label{rhosol}
	\Phi(x)=1+\sum_{I=0}^N\frac{\rho_I^2}{|x-x_I|^2}\,.
\end{equation}
This potential describes a multi-centered $N$-instanton configuration, where the $I^{\rm th}$ instanton is characterized by its size $\rho_I$ and position $x_I$. Despite being singular at the instanton location $x=x_I$, it leads to smooth, localized field strengths and gauge-invariant quantities. Notice that $N$ is an arbitrary number, and can be proved to represent the instanton number, relying on the singularity and asymptotic structure of the solution.

In non-Abelian ModMax theory, finding an analytic form for the potential $\Phi(x)$ is a rather difficult task due to the highly nonlinear nature of the BPS equations. To achieve this, we can address the problem perturbatively. To this end, we propose a power series expansion for the function $\Phi(x)$ in terms of the ModMax coefficient, namely%
\begin{equation}
\Phi(x)  =\Phi^{(0)}(x) +\gamma\,
\Phi^{(1)}(x)  +\gamma^{2}\,\Phi^{(2)}(x)  + \mathcal{O}\left(  \gamma^{3}\right)  \ .
\end{equation}
At the lowest order, the self-duality condition of ModMax theory [cf. Eq.~\eqref{BPS-cond}] is exactly that of Yang-Mills. Therefore, it implies the Laplace equation for $\Phi^{(0)}(x)$ in $\mathbb{R}^{4}$, whose solution is given in Eq.~\eqref{rhosol}.
At first order in $\gamma$, on the other hand, the self-duality condition implies a Poisson equation for $\Phi^{(1)}(x)$, which reads%
\begin{equation}
\nabla^{2}\Phi^{(1)}(x)  =F\left(  \Phi^{(0)}(x)  \right)  ^{1/2}\ ,\label{Lap1}
\end{equation}
where we have defined the source term as
\begin{equation}
	F\left(\Phi^{(0)}(x)\right)=\frac{1}{3(\Phi^{(0)})^2}\left[\big(\Phi^{(0)}\big)^2\Phi^{(0)}_{\mu\nu}\Phi^{(0)\mu\nu}  -4\Phi^{(0)}\Phi^{(0)\mu}\Phi^{(0)\nu}\Phi^{(0)}_{\mu\nu}+3\left(\Phi^{(0)}_\mu\Phi^{(0)\mu}\right)^2\right]\,,
\end{equation}
with $\Phi^{(0)}_\mu:=\nabla_\mu\Phi^{(0)}$ and $\Phi^{(0)}_{\mu\nu}:=\nabla_\mu\nabla_\nu\Phi^{(0)}$. Even though the source might look intricate, it is completely determined by the solution at the leading order $\Phi^{(0)}(x)$ [cf. Eq.~\eqref{rhosol}]. Thus, once such
harmonic function is fixed, then Eq.~\eqref{Lap1} can be solved analytically by
\begin{equation}\label{rho1sol}
\Phi^{(1)}\left(  x\right)  =-\int \dd ^{4}y \,G\left(  x-y\right)
F\left(\Phi^{(0)}(y)\right)^{1/2}\,, \;\;\;\; \mbox{where} \;\;\;\;\; G\left(  x-y\right)  =\frac{1}{4\pi^{2}}\frac{1}{\left\vert x-y\right\vert
^{2}}
\end{equation}
is the Green function of the $\mathbb{R}^{4}$ Laplacian. Formally, this is the general solution to first-order in the perturbative expansion of $\gamma$. 

For the one-instanton solution, 
we can use translation invariance to locate the solution at the origin of $\mathbb{R}^4$, 
leading to a very simple expression for the source, that is,
\begin{align}  
F\left(\Phi^{(0)}(x)\right)^{1/2}=\frac
{4\rho_0^2}{(t^{2}+x_i x^i+\rho_0^2)(t^{2}+x_i x^i)}\,.
\end{align}
A standard computation allows writing the solution in Eq.~\eqref{rho1sol} as the Fourier transform of a modified Bessel function of the second kind. Given the symmetry, the integral can be performed analytically, leading to
\begin{align}
\Phi^{(1)}(r) =- \ln\left(1+\frac{\rho_0^2}{r^2}\right) + \frac{\rho_0^2}{r^2}\,  \ln\left(1+\frac{r^2}{\rho_0^2}\right) \,,
\end{align}
where $r=|x|$. Consequently, the one-instanton solution, at first order in $\gamma$ reads
\begin{equation}
\Phi(r)  =1+\frac{\rho_0^2}{r^2}-\gamma\left[\ln\left(1+\frac{\rho_0^2}{r^2}\right) - \frac{\rho_0^2}{r^2}\,  \ln\left(1+\frac{r^2}{\rho_0^2}\right)\right]+\mathcal{O}\left(\gamma^2\right)\ .\label{rhoprimerorden}
\end{equation}
At the perturbative order we are working, the one-instanton solution constructed from the `t Hooft ansatz in Eq.~\eqref{tHansatz} with the solution~\eqref{rhoprimerorden} exactly reproduces the self-dual, one-instanton solution with $k=0$ of the previous section in \eqref{asol-k}, provided we identify
\begin{equation}
a(r)=-\frac{1}{2}\frac{r}{\Phi(r)}\frac{\dd\Phi(r)}{\dd r}=-\frac{\dd \ln\Phi(r)}{\dd \ln{r^2}}\ .
\end{equation}
The perturbative solution in Eq.~\eqref{rhoprimerorden} is remarkably simple, despite some intricate integrals appearing in the intermediate steps of the computation. This provides a strong indication that multi-instanton configurations do indeed exist at the first order in $\gamma$, and they reduce smoothly to the multi-instanton configurations of Yang-Mills theory when $\gamma$ vanishes. The profiles are obtained from Eq.~\eqref{rho1sol} by considering multi-instantons at lowest order, then performing the integral to obtain the first-order correction.

\section{Dirac spectrum on the Euclidean AdS boundary}\label{sec:Dirac}

The spectrum of the Dirac operator is necessary to obtain the boundary contribution to the topological invariants of the Pontryagin class. This is because, as explained in Sec.~\ref{sec: Instantons on constant curvature spaces}, the bulk integral of the Pontryagin density does not need to be an integer in spacetimes with negative curvature~\cite{Callan:1989em}. Besides the codimension-1 integral of the Chern-Simons form, the Dirac index receives a non-local contribution coming from the Atiyah-Patodi-Singer (APS) $\eta_D$-invariant, which measures the spectral asymmetry of the Dirac operator at the boundary~\cite{Atiyah:1975jf,Atiyah:1976jg,Atiyah:1976qjr}.

To obtain the spectrum, we project the Dirac operator in the fundamental representation of $SU(2)$ onto the conformal boundary of Euclidean AdS in the presence of non-Abelian ModMax fields, that is,
\begin{align}
    \slashed{D} = \gamma^i E^\mu_{\ i} \left(\partial_\mu + \frac{1}{4}\omega^{jk}{}_\mu \gamma_{[j}\gamma_{k]} + iA_\mu \right)\,,
\end{align}
where $i,j,k$ are $SO(3)$ internal indices. Here, $E_i=E^\mu_{\ i}\partial_\mu$ is the non-holonomic basis for $SO(3)$ vectors, whose dual $1$-form, the dreibein, is given by $e^i=e^i_{\ \mu} \diff{x^\mu}$. The latter is related to the metric of the conformal boundary of Euclidean AdS via $\bar{g}_{\mu\nu}=\delta_{ij}e^i_{\ \mu}e^j_{\ \nu}$. Indeed, from Eq.~\eqref{dsCCS} with $k=-1$, one can read $e^i=\sigma^i$, provided a suitable map between $SO(3)$ and $SU(2)$ indices as done in Sec.~\ref{sec:Multi-instantons} for the 't Hooft symbols. This identification greatly simplifies the computation. On the other hand, the spin connection $1$-form, defined as $\omega^{ij}=\omega^{ij}{}_\mu\dd x^\mu$, can be obtained in terms of the dreibein from the Cartan structure equation, i.e. $\dd e^i + \omega^{i}{}_{j}\wedge e^j=0$. The Dirac matrices, $\gamma^i$, span the Clifford algebra in Euclidean signature, namely, $\{\gamma_i,\gamma_j\}=2\delta_{ij}\mathbb{I}$. In three dimensions, they are nothing but the Pauli matrices, i.e. $\gamma^i = \tau^i$. Henceforth, we choose the following representation for the latter
\begin{align}
\gamma^1 &= \begin{pmatrix} & 0 & & 1 &  \\ & 1 & & 0 & \end{pmatrix}, & \gamma^2 &= \begin{pmatrix} & 0 & & -i & \\ & i & & 0 & \end{pmatrix}, & \gamma^3 &= \begin{pmatrix} & 1 & & 0 & \\ & 0 & & -1 & \end{pmatrix}.
\end{align}

To solve the eigenvalue problem, we define the self-adjoint operator $K_i=i\Sigma_i$ and the ladder operator $K_\pm = K_1\pm iK_2$, where $\Sigma_i$ are the dual vectors to the left-invariant forms of $SU(2)$, i.e. $\langle\sigma^i,\Sigma_j\rangle = \delta^i_j$. These operators satisfy the $SU(2)$ algebra, namely,
\begin{align}
    [K_i, K_j] &= i \epsilon_{ijk} K^k, & [K_3, K_\pm] &= \pm K_\pm, & [K_+, K_-] &= K_3.
\end{align}
Following Refs.~\cite{Hitchin:1974rbi,Bakas:2011nq,Canfora:2023bug}, here we consider the direct sum of unitary irreducible representations of $SU(2)$. Those states are be denoted by $|j,m\rangle$, with $j=0,1/2,1,3/2,\ldots$, including half-integer quantum numbers. The action of the self-adjoint operator on these states yields
\begin{align}
    K_\pm |j,m\rangle &= \sqrt{(j\mp m)(j\pm m +1)}\,|j,m\rangle\,, \\
    K_3 |j,m\rangle &= m\,|j,m\rangle\,.
\end{align}
Defining $\bar{K}_i = K_i - a_{0\pm}$, where $a_{0\pm}$ is given in Eq.~\eqref{apminfty}, and $\bar{K}_{\pm}=\bar{K}_1 \pm i \bar{K}_2$, the Dirac operator in the fundamental representation of $SU(2)$ can be written as
\begin{align}
    \slashed{D} = \left(\begin{matrix} \bar{K}_3 + \frac{3}{4} & \bar{K}_- \\
    \bar{K}_+ & -\bar{K}_3 + \frac{3}{4}  \end{matrix}
    \right)\,.
\end{align}
The eigenvalues of the Dirac operator are then obtained from the characteristic equation $\det(\slashed{D}-\lambda\mathbb{I})=0$. Thus, denoting $p\equiv j+m+1$ and $q\equiv j-m$, we obtain
\begin{align}
    \lambda_\pm &= \frac{1}{4} \pm 2\sqrt{\left(p-q-2a_{0\pm} \right)^2 + 4\left(2a_{0\pm} +pq \right) - 8a_{0\pm}\sqrt{pq}}\;. 
\end{align}
Then, the contribution of non-Abelian ModMax instantons to the spectrum of the Dirac operator becomes transparent. This provides an important step towards obtaining the spectral asymmetry of the Dirac operator at the conformal boundary of Euclidean AdS, which is crucial for computing the index of the Dirac operator analytically. If the latter is nonvanishing, it would indicate the presence of axial anomaly induced by self-dual instantons in non-Abelian ModMax theory.

\section{Einstein gravity and nonlinear conformal matter}\label{sec:Self-gravitating}

To study the backreaction of non-Abelian ModMax fields when coupled to gravity, we consider a more general case that includes a conformally coupled scalar field with a self-interacting quartic potential, maintaining as a guiding principle the conformal invariance of the matter sector under $g_{\mu\nu}\to\Omega^2g_{\mu\nu}$, $A_\mu\to A_\mu$, and $\phi\to\Omega^{-1}\phi$, with $\Omega=\Omega(x)$ an arbitrary, analytic function of the coordinates. For a constant scalar field, the theory under consideration reduces to Einstein gravity minimally coupled to non-Abelian ModMax fields. In particular, the action principle we consider is
\begin{equation}
\label{eq:GRplusnAMM}
I[g,A,\phi]=\int_\mathcal{M}\dd^4x\sqrt{|g|}\Big[\kappa (R - 2\Lambda)- \dfrac{1}{2}\nabla^{\mu}\phi\nabla_{\mu}\phi - \dfrac{1}{12}R\phi^2 -\nu\phi^4\Big] + I_{\rm NAM}\,,
\end{equation}
where $\kappa=(16\pi G)^{-1}$ denotes the gravitational constant, and $I_{\rm NAM}$ is given in Eq.~\eqref{nAMM}. The field equations for the non-Abelian ModMax fields are given in Eq.~\eqref{eom-A}. Extremizing the action with respect to the metric and the scalar field yield
\begin{subequations}\label{eom-g}
    \begin{align}
    \mathcal{E}_{\mu\nu} &:= G_{\mu\nu} + \Lambda g_{\mu\nu}-\dfrac{1}{2\kappa}T^{(\phi)}_{\mu\nu}-\frac{1}{2\kappa}T_{\mu\nu}^{(A)}=0 \,, \label{eom-mm-sf}\\ 
    \mathcal{E}_{(\phi)}&:= \square \phi - \dfrac{1}{6}R\phi - 4\nu\phi^3 = 0 \,,\label{eom-p}
\end{align} 
\end{subequations}
respectively, with $G_{\mu\nu}=R_{\mu\nu}-\tfrac{1}{2}g_{\mu\nu}R$ being the Einstein tensor. Additionally, $T^{(A)}_{\mu\nu}$ is the non-Abelian ModMax energy-momentum tensor given in Eq.~\eqref{Tmunu}, and its scalar counterpart is defined by
\begin{equation}\label{Tmunu-scalar-field}
T_{\mu\nu}^{(\phi)}=\nabla_\mu\phi\nabla_\nu\phi -\frac{1}{2}g_{\mu\nu }\nabla_\alpha\phi\nabla^\alpha\phi +\frac{1}{6}(g_{\mu\nu }\Box-\nabla_\mu\nabla_\nu +G_{\mu\nu })\phi^2-\nu g_{\mu\nu}\phi^4\,.
\end{equation}
The trace of the scalar stress tensor vanishes on shell, as it can be written as $g^{\mu\nu}T_{\mu\nu}=\phi\mathcal{E}_{(\phi)}$, where the scalar field equation is defined in~\eqref{eom-p}. Thus, due to the conformal invariance of the full matter sector, solutions to the field equations must have constant Ricci scalar, that is, $R=4\Lambda$. 

\subsection{Gravitating non-Abelian dyons}

\begin{figure}[h]
\centering
\includegraphics[scale=0.55]{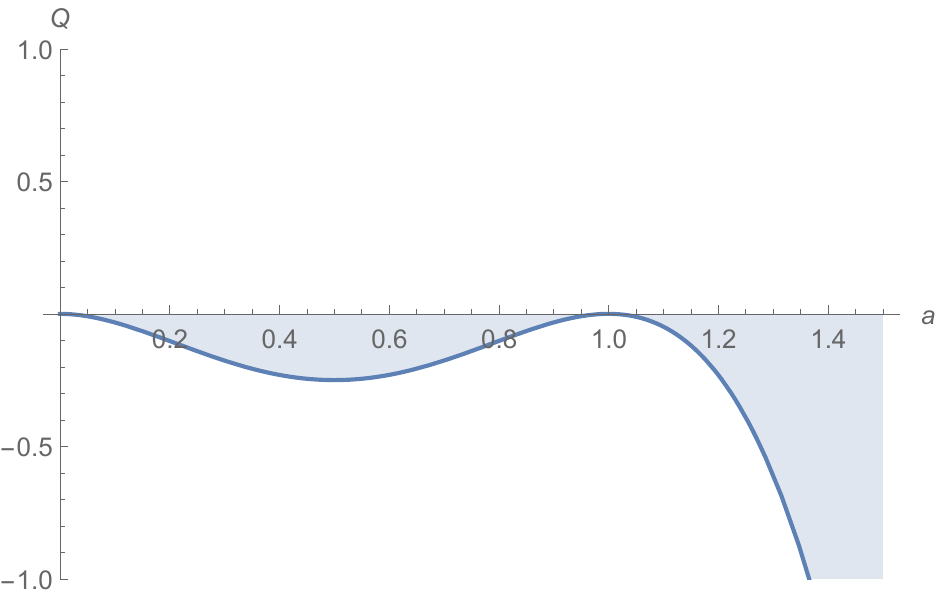}
    \caption{Plot of the potential $V(a)$ as a function of the generalized coordinate $a$. The shaded area represents the allowed region for solutions, where $Q\geq V(a)$.}
    \label{QvsA}
    \end{figure}

For the non-Abelian field, we consider the same ansatz as in Eq. \eqref{A}, while for the metric, we choose a conformally flat space 
\begin{equation}\label{ds-WH}
	\dd s^2=f^2(r)\left(\dd r^2+\frac{1}{4}(\sigma_1^2+\sigma_2^2+\sigma_3^2)\right)\,.
\end{equation}
Since our aim is to look for self-gravitating solutions, we shall not impose the generalized (anti-)self-duality condition, as the stress-energy tensor would vanish identically by virtue of the identity~\eqref{TmunuSD}. Instead, we insert the ansatz~\eqref{A} directly into the field equations~\eqref{eom-A}, reducing the system to a single second-order differential equation for the function $a(r)$, i.e.,
\begin{equation}\label{DE-a}
	a''-4e^{2\gamma}a(a-1)(2a-1) =0\,,
\end{equation}
which admits a first integral, resulting in
\begin{equation}\label{dadr-WH}
a'^2 - 4e^{2\gamma}a^2(a-1)^2 = Q\,,
\end{equation}
where $Q$ is an integration constant. Solutions of the non-Abelian ModMax equation can be analyzed by rewriting Eq.~\eqref{dadr-WH} in terms of a potential for the non-Abelian profile, namely, $a'^2 + V(a) = Q$, where $V(a):=-4e^{2\gamma}a^2(a-1)^2$, whose behavior is given in Figure~\ref{QvsA}.
    
The potential possesses a well, which is characteristic of periodic solutions, and a second branch with unbounded trajectories. There are three extrema values, located at $a=0$, $a=1$, and $a=1/2$. While the first two are pure gauge solutions, the latter corresponds to a constant solution that is not pure gauge, that represents a meronic solution of the non-Abelian ModMax theory, which generalizes that found in Yang-Mills theory~\cite{deAlfaro:1976qet}.

\begin{figure}[h]
    \centering
    \includegraphics[scale=0.5]{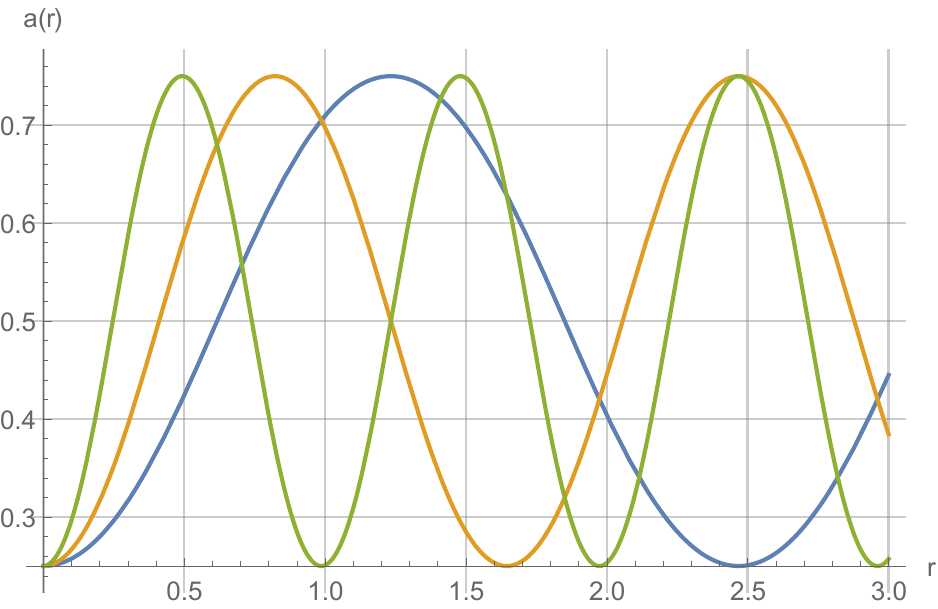}
    \caption{Profile of the oscillatory solution of the ModMax field equations. Here, we choose the initial conditions $a(0)=0.25$, $a'(0)=0$, and we have used $\gamma = \{ \ln 2\,, \ln 3\,, \ln 5\}$ for the blue, orange, and green plots, respectively. Notice that the role of the ModMax parameter $\gamma$ is directly related to the frequency of the oscillatory solution: the larger its value, the higher the frequency.}
    \label{asol-WH}
\end{figure}

We deal with the second-order equation~\eqref{DE-a} employing numerical techniques, subject to the initial conditions $a(0)=a_0$, with $0<a_0<1$, and $a'(0)=0$ for simplicity. The solution for different values of $\gamma$ is presented in Figure~\ref{asol-WH}.
Notice that the frequency of the oscillatory solution increases with $\gamma$.

The stress-energy tensor of the self-gravitating non-Abelian ModMax fields [cf. Eq.~\eqref{Tmunu}] can be written as an Euclidean perfect fluid, that is,
\begin{align}
    T_{\mu\nu} = -\rho(r)u_\mu u_\nu + p(r)h_{\mu\nu}\,,
\end{align}
where we have defined $u=u^\mu\partial_\mu=\tfrac{1}{f(r)}\partial_r$ as the Euclidean four-velocity of a fluid element satisfying $u_\mu u^\mu=1$, while $h_{\mu\nu}=g_{\mu\nu}-u_\mu u_\nu$ is the projector onto the transverse components of the fluid. Conformal invariance of the non-Abelian ModMax fields implies that its equation of state is $p(r)=\rho(r)/3$, being an ultra-relativistic fluid. One can check that it has zero acceleration, vorticity, and shear, while its energy density and expansion are
\begin{align}
   \rho(r) = \frac{3Qe^{-\gamma}}{f^4(r)} \;\;\;\;\; \mbox{and} \;\;\;\;\; \Theta(r) := \nabla_\mu u^\mu = \frac{3f'(r)}{f^2(r)}\,.
\end{align}
respectively. Since both quantities depend on the metric function, and the non-Abelian ModMax equation has already been solved, the Einstein field equations will determine their behavior. Before doing so, let us first analyze non-Abelian gauge invariant charges.

\subsection{Gauge invariant charges}

In nonlinear electrodynamics, the electric and magnetic charges are defined by integrating the constitutive relations and the Abelian field strength $2$-forms, respectively, at spacelike infinity, as in Maxwell theory.

However, it is well known that these quantities cease to be gauge invariant when the fields are promoted to take values in a non-Abelian group. In Ref.~\cite{Corichi:2000dm}, a gauge-invariant definition of charges in Einstein-Yang-Mills theory was proposed. The latter can be regarded as the non-Abelian generalizations of the standard electric and magnetic charges in Maxwell theory. In this approach, the non-Abelian electric and magnetic charges are defined as~\cite{Corichi:2000dm}
\begin{equation}\label{Qs-Ps}
	\mathcal{Q}_e:=\frac{1}{4\pi }\oint_{\mathbb{S}^2}|\tilde{P}|\,\quad \text{and}\quad \mathcal{Q}_m:=\frac{1}{4\pi }\oint_{\mathbb{S}^2}|F|\,,
\end{equation}
respectively, where the gauge-invariant $2$-forms are given by 
\begin{subequations}
\begin{align}
|\tilde{P}|_{\alpha\beta}&=\sqrt{\big|2\Tr(\tilde{P}^{\mu\nu}\varepsilon_{\mu\nu}\tilde{P}^{\lambda\rho}\varepsilon_{\lambda\rho})\big|}\varepsilon_{\alpha\beta}\,, \\ |F|_{\alpha\beta}&=\sqrt{\big|2\Tr(F^{\mu\nu}\varepsilon_{\mu\nu}F^{\lambda\rho}\varepsilon_{\lambda\rho})\big|}\varepsilon_{\alpha\beta}\,.
\end{align}
\end{subequations}
Here, $\varepsilon_{\mu\nu}$ represents the area element of $\mathbb{S}^2$ and, in Euclidean signature, it is normalized such that $\varepsilon_{\mu\nu}\varepsilon^{\mu\nu}=2$. 

Evaluating the non-Abelian charges~\eqref{Qs-Ps} on the $\mathfrak{su}(2)$-valued gauge field in Eq.~\eqref{A}, we find that they can be expressed as $\mathcal{Q}_e=-a'e^{-\gamma}$ and $ \mathcal{Q}_m=2a(a-1)$. As the non-Abelian electric and magnetic charges are nontrivial for the solution studied here, we interpret this configuration as a gravitating non-Abelian dyon.  

From the solution in Eq.~\eqref{dadr-WH}, we find that these charges satisfy
\begin{equation}\label{Qe-Qm=Q}
	\mathcal{Q}_e^2-\mathcal{Q}_m^2=Qe^{-2\gamma}\,.
\end{equation}
Interestingly, similar to the Abelian case, the backreaction of the ModMax fields is proportional to the sum (difference) of the electric and magnetic charges squared in Lorentzian (Euclidean) signature. Even more, when both electric and magnetic non-Abelian charges satisfy $\mathcal{Q}_e=\pm \mathcal{Q}_m$, the non-Abelian ModMax fields become (anti-)self-dual, producing no backreaction whatsoever. Since we are interested in self-gravitating solutions, we will not assume the self-duality condition from hereon.

\subsection{Euclidean AdS wormholes and gravitational instantons}

Euclidean wormholes are defined as regular geometries with nowhere vanishing area coordinate that connect different asymptotic regions, with or without boundaries. These configurations have recently attracted significant interest in the presence of negative cosmological constant since, in the context of the AdS/CFT correspondence~\cite{Maldacena:1997re,Witten:1998qj,Gubser:1998bc}, they provide a framework for studying holography with multiple boundaries~\cite{Witten:1999xp,Maldacena:2004rf,Bergshoeff:2005zf,Arkani-Hamed:2007cpn,Skenderis:2009ju}. Moreover, it has been understood that such geometries encode nonperturbative effects that play an important role in black hole physics. For instance, they offer a partial resolution of the black hole information paradox~\cite{Saad:2018bqo} and they are essential for understanding the Ryu-Takayanagi proposal for holographic entanglement entropy~\cite{Ryu:2006bv,Ryu:2006ef}, as explored in~\cite{Jensen:2014lua,Maxfield:2014kra,Balasubramanian:2014hda}.

Despite their Euclidean nature, they exhibit a wide range of applications in both particle physics and cosmology~\cite{Hebecker:2018ofv}. They also give rise to several conceptual challenges, such as the factorization problem in holography~\cite{Maldacena:2004rf,Arkani-Hamed:2007cpn,Bergshoeff:2004pg,Andriolo:2022rxc,Jonas:2023ipa,Gutperle:2002km}. A broad variety of such configurations have been explored in different settings, including supergravity theories~\cite{Hertog:2017owm,Anabalon:2018rzq,Anabalon:2020loe,Anabalon:2023kcp,Astesiano:2023iql}, higher-curvature extensions of gravity~\cite{Dotti:2006cp,Dotti:2007az,Dotti:2010bw}, and in the presence of matter fields such as Yang-Mills fields~\cite{Maldacena:2004rf,Hosoya:1989zn}, phantom fields~\cite{Blazquez-Salcedo:2020nsa,Chew:2021vxh}, Higgs fields~\cite{Betzios:2024zhf}, Skyrmions~\cite{Canfora:2025roy}, and with scalar fields~\cite{Anabalon:2012tu,Cisterna:2021xxq,Barrientos:2022avi,Cisterna:2023uqf}. 
In what follows, we construct Euclidean AdS wormholes in the presence of non-Abelian ModMax dyons and conformally coupled scalar fields.

To look for Euclidean wormhole solutions in Einstein-non-Abelian-ModMax, we perform a radial diffeomorphism such that the metric in Eq.~\eqref{ds-WH} can be written as
\begin{align}
    \dd s^2 = \dd r^2 + \frac{h(r)}{4}\left(\sigma_1^2+\sigma_2^2 + \sigma_3^2 \right)\,,
\end{align}
where $h(r)$ is referred to as the area coordinate. Since the trace of the energy-momentum tensor vanishes on shell by virtue of conformal invariance of scalar and gauge fields, the field equations for the metric yields $R=4\Lambda$. This condition, when evaluated on the ansatz~\eqref{ds-WH}, gives the second-order nonlinear equation
\begin{align}
\label{eq:metricf-Lambda}
   -\frac{3h''}{h} + \frac{6}{h} = 4\Lambda\,.
\end{align}
This equation can be solved analytically, and its solution is 
\begin{equation}\label{hsol-Lneq0}
	h(r) = \frac{3}{2\Lambda} + \frac{\rho_0^2}{2}\cos\left(2\sqrt{\frac{\Lambda}{3}}\,(r-r_0)  \right) \,,
\end{equation}
where $\rho_0$ and $r_0$ are integration constants of dimension $1$, and the cosmological constant is nonvanishing. In the de Sitter case, the solution is positive definite $\forall r\in\mathbb{R}$ in the range $\rho_0\in(-\ell_{\rm dS},\ell_{\rm dS})$, where $\ell_{\rm dS}$ is the de Sitter radius defined as $\Lambda=3/\ell_{\rm dS}^2$. In this case, although the solution is regular, the area coordinate exhibits oscillatory behavior rather than a monotonic increase. Therefore, we interpret this solution as a one-parameter extension of the gravitational instanton found in Ref.~\cite{Hosoya:1989zn}.

In the negative cosmological constant case, on the other hand, the solution describes an Euclidean AdS wormhole as long as $\rho_0>\ell_{\rm AdS}$, with $\ell_{\rm AdS}$ being the AdS radius defined as $\Lambda=-3/\ell_{\rm AdS}^2$. In such a case, the throat is located at $r=r_0$ where the area coordinate is minimum, and its value is given by $h(r_0) = (\rho_0^2-\ell_{\rm AdS}^2)/2>0$. 

If the cosmological constant vanishes, then Eq.~\eqref{eq:metricf-Lambda} must be solved from scratch, whose solution is given by 
\begin{align}\label{hsol-Leq0}
    h(r) = r^2 + r_0^2 + c_0\, r  \,,
\end{align}
where $c_0$ and $r_0$ are integration constants. Notice that, if $c_0>0$, this solution is also nowhere vanishing $\forall r\in\mathbb{R}$. Indeed, if $c_0=0$, the area coordinate is still positive definite as long as $r_0\neq0$. This solution also represents a wormhole with a vanishing cosmological constant that connects two asymptotically flat regions, with a throat located at $r=0$, where the area coordinate is a minimum.

Let us analyze the remaining components of the field equation for the metric. For a nonvanishing cosmological constant, we find two disconnected branches of solutions. In the first one, the scalar field is constant, say $\phi=\phi_0$, and the following conditions must hold 
\begin{align}
    \Lambda = -6\nu\phi_0^2\,, \qquad \mbox{and} \qquad Q=-\frac{e^\gamma (\phi_0^2-12\kappa)(4\nu^2\phi_0^4\rho_0^4-1)}{48\nu\phi_0^2}\,.
\end{align}
In this case, the theory reduces to Einstein gravity minimally coupled to non-Abelian ModMax fields, with a constant scalar field that induces a shift in the gravitational and cosmological constants. 

The other branch, however, admits a nontrivial scalar field that supports the gravitational configuration, that is,
\begin{align}
\label{eq:scalarfield-sol}
    \phi(r)= \frac{1}{2\sqrt{-\nu h(r)}}\,.
\end{align}
Reality conditions imply that $\nu<0$. Since we are restricting ourselves to the case where the area coordinate is positive definite and nowhere vanishing, the scalar field is everywhere regular. Additionally, the remaining components of the field equations lead to a modified relation between $Q$ and $\rho_0$ due to the presence of the self-gravitating scalar field, that is, 

\begin{equation}\label{Q-Lneq0}
	Q=\frac{(72\kappa\nu+\Lambda)e^\gamma}{48\nu\Lambda}-\frac{\Lambda\kappa\rho_0^4 e^\gamma}{6}\,.
\end{equation}
Notice that the presence of the self-interacting nonlinear scalar potential is crucial for this solution to exist, and it implies that the limit $\nu\to0$ is not smooth. Additionally, this solution is compatible with the (anti-)self-dual limit, that is, $Q\to0$, where the non-Abelian electric and magnetic charges become equivalent [cf.~Eq.~\eqref{Qe-Qm=Q}]. In that case, the non-Abelian gauge field are aligned with the left-invariant one-forms of $SU(2)$, as in Eq.~\eqref{A}. 

If the cosmological constant vanishes, the constraint among the parameters and the couplings of the theory reads
\begin{equation}\label{Q-Leq0}
	Q=\frac{e^\gamma}{48\nu }-\frac{\kappa(c_0^2-4r_0^2)e^\gamma}{2}\,.
\end{equation}
The conditions~\eqref{Q-Lneq0} and~\eqref{Q-Leq0} do not affect the previously discussed constraints on the integration constants, ensuring that $h(r)$ remains positive definite.

Finally, it is worth mentioning that, the scalar field, the ModMax fields, and the metric are all completely regular. For instance, the $SU(2)$ invariants for the solution in Eq.~\eqref{dadr-WH} are given by 
\begin{subequations}
	\begin{align}
		X&= -\frac{12a^2(a-1)^2(e^{2\gamma}+1)+3Q}{h^2}\,, \\ Y&= -\frac{12a(a-1)\sqrt{4e^{2\gamma}a^2(a-1)^2+Q}}{h^2}\,.
	\end{align}
\end{subequations}
with $a=a(r)$ having the oscillatory behavior displayed in Figure~\ref{asol-WH}. Thus, these invariants are regular $\forall r\in\mathbb{R}$. Additionally, from the Kretschmann invariant, we can check that 
\begin{align}
     R^{\mu\nu}_{\lambda\rho}R_{\mu\nu}^{\lambda\rho} &= 
    \begin{cases}
        \displaystyle \;\; \frac{8\Lambda^2}{3}+\frac{(\Lambda^2\rho_0^4-9)^2}{6\Lambda^2h^4}
        & \text{for } \;\; \Lambda\neq 0, \\[1.2em]
        \displaystyle \;\; \frac{3(c_0^2-4r_0^2)^2}{2h^4}
        & \text{for } \;\; \Lambda=0 \,.
    \end{cases}
\end{align}
where $h=h(r)$ is given in Eqs.~\eqref{hsol-Lneq0} and~\eqref{hsol-Leq0}, depending on whether the cosmological constant vanishes or not. In any case, these functions never vanish, provided the conditions discussed above are satisfied. One can prove that other curvature invariants behave similarly. Therefore, these conditions guarantee that the metric, scalar, and non-Abelian ModMax dyons are fully regular, showing that nonlinear effects are relevant for avoiding singularities.  

\section{Discussion}\label{sec:Discussion}

In this work, we analyzed the non-Abelian extension of ModMax electrodynamics. This theory is characterized by a single dimensionless parameter, and it reduces to Yang-Mills theory in the limit when such a parameter vanishes. In particular, the Euclidean case with internal gauge group $SU(2)$ was studied in detail. 

We showed that the field equations admit (anti-)self-dual instantons on constant-curvature backgrounds, which render the energy-momentum tensor identically zero, much like in the Yang-Mills case. These configurations generalize the BPST instanton~\cite{Belavin:1975fg} on flat space and the ILP instanton~\cite{Ivanova:2017wun} on Euclidean de Sitter and anti-de Sitter spaces. In the latter case, the non-Abelian profile depends asymptotically on the instanton size, and therefore its Chern-Pontryagin index is not an integer, similar to what happens with Yang-Mills instantons on negative-curvature backgrounds, as shown by Callan and Wilczek in Ref.~\cite{Callan:1989em}. We obtained the spectrum of the Dirac operator which is crucial for determining the boundary contributions to topological invariants of the Pontryagin class.

Then, the possibility of constructing multi-instanton configurations in non-Abelian ModMax theory was discussed, a prospect suggested by the theory's scale invariance. Due to the strong nonlinearity of the BPS equations, obtaining analytic solutions is extremely challenging. For this reason, we analyzed the system perturbatively in the dimensionless parameter. In this perturbative approach, we succeeded in constructing a solution for an $N$-instanton configuration to first order in the parameter expansion. Finally, we considered the non-Abelian ModMax theory coupled to General Relativity with a cosmological constant and conformally coupled scalar fields. Owing to the conformal invariance of the matter sector, we found new analytic self-gravitating solutions of this system, which can be interpreted as Euclidean wormholes and gravitational instantons with and without the cosmological constant. Some of their properties were analyzed, including the gauge-invariant charges associated with the non-Abelian fields, the role of the ModMax parameter in shaping the gauge profile and the covariant non-Abelian charges, and the interpretation of the ModMax energy-momentum tensor as that of an ultra-relativistic fluid with nontrivial expansion, among others.

Interesting questions remain open. For instance, it would be valuable to study holographic aspects of these configurations on Euclidean AdS space. This could be relevant for analyzing the non-Abelian magnetotransport dual conformal field theories. Additionally, the spectrum of the Dirac operator on the boundary of Euclidean AdS can be used to determine the parity anomaly on the dual CFT by computing the spectral asymmetry via the Atiyah-Patodi-Singer invariant~\cite{Alvarez-Gaume:1984zst}. These questions are left for future investigation.

\begin{acknowledgments}
    The authors thank Luis Avilés, Eloy Ayón-Beato, and Daniel Flores-Alfonso for fruitful discussions.
    This work is partially funded by the Agencia Nacional de Investigación y Desarrollo (ANID) through Fondecyt Regular grants No. 1240043, 1240048, 1251523, 1252053, 1221504, and 1230112, as well as by ANID Proyecto de Exploración 1325001.
\end{acknowledgments}

\bibliographystyle{JHEP}
\bibliography{biblio.bib}

\end{document}